\documentclass[twocolumn,showpacs,preprintnumbers,amsmath,amssymb]{revtex4}
\usepackage{graphicx}% Include figure files
\usepackage{dcolumn}% Align table columns on decimal point
\usepackage{bm}% bold math

\begin{document}

%\preprint{APS/123-QED}

\title{Measurement of Optical Response of a \\
Detuned Resonant Sideband Extraction Interferometer \\ 
LIGO-P060007-00-R}% Force line breaks with \\

\author{Osamu Miyakawa, Robert Ward, Rana Adhikari, Matthew Evans, Benjamin Abbott, Rolf Bork, Daniel Busby, Jay Heefner, Alexander Ivanov, Michael Smith, Robert Taylor, Stephen Vass, and Alan Weinstein}
%\author{Monica Varvella}%
\affiliation{%
LIGO Laboratory, California Institute of Technology, Pasadena, CA 91125, USA
}%
\author{Monica Varvella}
\affiliation{
Laboratoire de l'Accelerateur Lineaire Universite' Paris Sud, Orsay cedex, 91898, France
}%
\author{Seiji Kawamura}
\affiliation{
National Astronomical Observatory of Japan, Tokyo 181-8588, Japan
}%
\author{Fumiko Kawazoe and Shihori Sakata}
\affiliation{
Ochanomizu University, Otsuka, Bunkyo-ku, Tokyo, 112-8610, Japan
}%
\author{Conor Mow-Lowry}
\affiliation{
Centre for Gravitational Physics, 
The Australian National University, Acton, ACT 0200, Australia
}%

\date{\today}% It is always \today, today,

\begin{abstract}
We report on the optical response of
% development of a lock acquisition and control scheme for 
a suspended-mass detuned
resonant sideband extraction (RSE) interferometer with power
recycling. 
The purpose of the detuned RSE configuration is to manipulate and
optimize the optical response of the interferometer to differential
displacements (induced by gravitational waves)
as a function of frequency, independently of other
parameters of the interferometer.
The design of our interferometer results in an
optical gain with two peaks: an RSE optical resonance at around 4 kHz 
and a radiation pressure induced optical spring at around 41 Hz. 
We have developed a reliable procedure for acquiring lock and
establishing the desired optical configuration. 
In this configuration, we have measured the
optical response to differential displacement and found good agreement
with predictions at both resonances and all other relevant frequencies.
These results build confidence in both the 
theory and practical implementation of the 
more complex optical configuration being planned for Advanced LIGO.

\iffalse  % older version
Simultaneously optical gain in the differential mode of
arms was measured. The measured optical gain consists of two resonant
peaks. One is a radiation pressure induced optical spring around
41\,Hz and another one is an optical resonance around 4\,kHz due to
optical response of the RSE interferometer. Both peaks are caused by a
detuned signal recycling cavity which changes the phase of carrier for
each arm cavity. %The correlation between these two resonances has a
possibility to circumvents the standard quantum limit\,(SQL) in the
next generation gravitational wave detectors.
\fi

\end{abstract}


%Use showkeys class option if keyword
\maketitle

%\section{Introduction}

Currently the first generation of ground-based laser interferometric
gravitational wave (GW) observatories, including LIGO~\cite{Sigg02},
VIRGO~\cite{Fiore02}, GEO~\cite{Willke02} and TAMA~\cite{Ando02}, are in
operation.
Together, they form a global network for the detection and study
of GWs and their astrophysical sources.
However, 
%the expected event rate for detectable
%GWs is still very low with the present sensitivity, and 
more sensitive detectors are required 
in order to detect significant numbers of sources.
Advanced LIGO~\cite{M990288,M990080} is one
of the next-generation of planned gravitational wave detectors 
which currently plans to employ a detuned signal mirror 
in order to manipulate and optimize the frequency response of the detector.
Such configurations hold the possibility of circumventing
~\cite{Buonanno01a,Buonanno01b,Buonanno02} 
the free-mass Standard Quantum Limit (SQL)
on the measurement of small displacements~\cite{Braginsky68}
at frequencies between those where shot noise dominates
and those where radiation pressure noise dominates.

Long-baseline gravitational wave detectors
are based on Michelson interferometers,
which are designed to be sensitive to small 
differential displacements of the arms.
The Initial LIGO, VIRGO, and TAMA300 detectors
add Fabry-Perot resonant optical cavities in the arms,
and a power recycling cavity formed from a mirror
between the laser and the Michelson,
to enhance the optical gain of the 
displacement measurement.
The next generation of detectors aims to improve on
the sensitivity by making use of higher-powered lasers
and more complex optical configurations.

\iffalse  % save this historical review for a longer paper!
The concept of the SQL in precision measurement was disscussed by
Braginsky~\cite{Braginsky68} in 1968. The two-photon formalism which
explains the quantum noise in the interferometric gravitational wave
detector using the concept of input vacuum from the signal port was
proposed by Cave and Schumaker~\cite{Cave85,Schumaker85} in
1981. Various ideas for circumventing the SQL using different optical
configurations were proposed, such as speed meter, squeezed vacuum,
and filter cavities.
% but all these ideas requires special
% optical configurations or special detection systems, and they are not
% compatible to the conventional large-scale interferometers.
\fi

Signal recycling (SR)
was proposed by Meers~\cite{Meers88} as a means 
to reduce the shot noise contribution to 
the displacement signal over a narrow band near DC.
By using a signal mirror at the asymmetric port
of the Michelson to form a signal recycling cavity,
the storage time of the signal sidebands can be increased,
leading to enhanced sensitivity near DC but
reduced sensitivity at higher frequencies;
the detector bandwidth is reduced.
Thus, it is most effective to use this technique
with interferometers whose arms are optical delay-lines (such as GEO)
or low-finesse cavities. High-gain power recycling (PR)
is used to enhance the overall optical gain,
resulting in a configuration referred to as dual recycling.

Mizuno~\cite{Mizuno93} proposed a complementary configuration,
called resonant sideband extraction (RSE).
In this configuration,
the interferometer arms are high finesse cavities,
the power recycling gain is correspondingly lower,
and a signal extraction cavity 
is used to resonantly extract the signal sidebands from the arms,
thereby enhancing the signal bandwidth.
The finesse of the 
relatively low-loss arm cavities can be made large,
with RSE compensating for the resulting loss of bandwidth.
The required optical gain of the
higher-loss power recycling cavity 
is correspondingly reduced,
resulting in both higher overall optical gain,
and much reduced requirements for compensation of the thermal load
on the transmissive optics in the power recycling cavity
(beam splitter and arm input test masses).
As interferometric detectors move to higher
laser power, this thermal loading can be a severe problem.
Thus, Advanced LIGO has chosen as its baseline optical configuration
a power-recycled Michelson interferometer
with high-finesse Fabry-Perot arms and 
an RSE signal extraction cavity~\cite{Strain03}.

When all other noise sources are reduced,
the sensitivity of such an interferometer is limited
by quantum sensing noise, which enters the interferometer through
the laser light incident at the symmetric port
and the vacuum incident at the asymmetric port of the Michelson
(where the gravitational wave signal exits)~\cite{Cave85,Schumaker85}.
At high frequencies, photon quantum shot noise
dominates; this contribution to the sensitivity
can be reduced by increasing the laser power.
At low frequencies, quantum fluctuations in the radiation pressure on
the test masses dominate; this contribution
can be reduced by decreasing
the laser power. 
Ref.~\cite{Buonanno01a,Buonanno01b} 
derives a description of these quantum-limited noise
sources with their correlations combined consistently.
In the presence of additional noise sources
such as seismic noise and thermal noise 
in the test masses or their suspensions,
the laser power 
can be chosen to optimize sensitivity.
In addition, 
by choosing the phase advance of the carrier laser light
in a signal cavity (the ``tune''),
the frequency dependence of the
combined quantum contribution to the sensitivity can be manipulated
between the complementary regimes of signal recycling and
resonant sideband extraction.
Such systems are conventionally referred to as ``detuned''.

A detuned signal cavity will exhibit optical resonances
at frequencies that can be chosen to optimize
sensitivity in the presence of other noise sources.
A peak at higher frequency arises from the unbalanced response 
of a GW sideband resonating in the detuned SR cavity
(the RSE optical resonance).
A peak at lower frequency arises because the GW sidebands
induced by the differential displacement of arm cavities
enter a signal cavity detuned from resonance,
forming a radiation pressure induced opto-mechanical spring
which will enhance the optical response at the spring's 
resonant frequency.

%\section{Optical response of detuned RSE}

\begin{figure}[tbp]
\includegraphics[width=7.5cm]{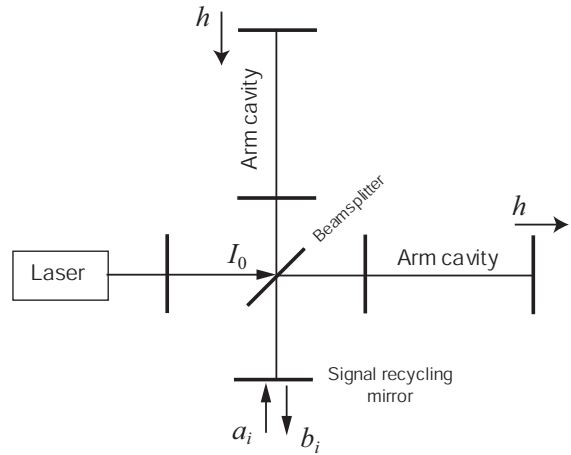}% Here is how to import EPS art
\caption{\label{fig:rse_config}
Schematic view of an RSE interferometer. 
$a_i$ and $b_i$ are the vacuum input and signal output. 
$h$ is the signal field due to the differential motion of the arm lengths
induced by a gravitational wave strain.
$I_0$ is the input laser power at the beamsplitter,
enhanced by the power recycling cavity.
%{\bf ((Fix the figure! $D_i \to I_0.$))}
}
\end{figure}

The optical response and noise spectra 
of the detuned RSE optical configuration
was analyzed by Buonnano and Chen~\cite{Buonanno01b} using the KLMTV
formalism~\cite{Kimble02}. 
Figure~\ref{fig:rse_config} shows the
relationship between the input vacuum field $a_i$ 
(carrying the quantum noise) to the signal port output
field $b_i$, the input laser power $I_0$, and the 
gravitational wave strain signal $h$. 
These quantities are related by Eq.(2.26) in~\cite{Buonanno01b}. 
The relation between $a$ and $h$ 
determines the quantum-limited strain sensitivity for the interferometer. 
The sensitivity as a function of frequency exhibits 
two dips in between the frequency regions where
radiation pressure noise and the shot noise dominate,
corresponding to the optical spring and detuned RSE optical resonances.
Buonanno and Chen~\cite{Buonanno01a,Buonanno01b,Buonanno02}
showed that in this configuration,
``ponderomotive squeezing'' will induce
correlations in the quantum noise.
The possibility exists of achieving quantum-limited sensitivity
beyond the ``standard'' quantum limit
in the frequency region around these two resonances.

To measure the  optical gain of the interferometer response
to a strain $h$, the end test masses are displaced differentially
by application of an external sinusoidal force. 
Typically, the applied displacement is large,
so that $h \gg a_i$ and the input vacuum $a_i$ is ignored.
In that approximation, the ratio between $h$ and $b_i$ in Eq.(2.26) of
~\cite{Buonanno01b} is the optical gain,
\begin{equation}
% \left|\frac{b_\zeta}{h}\right|=\sqrt{2\kappa}\tau \cdot \left|
% \frac{D_1 \sin \zeta+D_2 \cos \zeta}{M \cdot h_\mathrm{SQL}} \right|,
\frac{b_\zeta}{h} = 
  \frac{\sqrt{2\kappa}\tau e^{i\beta} }{M \cdot h_\mathrm{SQL}} 
   \left( D_1 \sin \zeta+D_2 \cos \zeta \right) ,
\label{eq:tf}
\end{equation}
where $b_\zeta$ is the output field with readout phase $\zeta$. 
$M$ and $D_i$ are defined as:
\begin{eqnarray}
M&=&1+\rho^2 e^{4i\beta}-2\rho e^{2i\beta}
    \left(\cos 2\phi+\frac{\kappa}{2}\sin 2\phi\right), \\
D_1&=&-(1+\rho e^{2i\beta})\sin \phi,\,D_2=(1-\rho e^{2i\beta})\cos \phi.
\end{eqnarray}
In these equations,
$2\beta=2\, \mathrm{atan} \Omega/\gamma$ is the net phase gained
by the laser light due to a sinusoidal 
GW with angular frequency $\Omega$ in the arm cavity, $\gamma=Tc/4L$
is the half bandwidth of the arm cavity, $T$ is the power
transmissivity of the arm cavity input mirrors, $L$ is the length
of the arm cavity, $\tau$ is the amplitude transmissivity 
and $\rho$ is the amplitude reflectivity of the SR mirror,
and $\phi$ is the detuning of the signal cavity from carrier resonance.
% (we approximate the SR mirror to have negligible loss;
% $\tau^2+\rho^2=1$).
$\kappa$ is an effective coupling constant which
relates the mirror motion to the output signal,
\begin{equation}
\kappa=\frac{2(I_0/I_\mathrm{SQL})\gamma^4}{\Omega^2(\gamma^2+\Omega^2)},
\quad
I_\mathrm{SQL}=\frac{mL^2\gamma^4}{4\omega_0},
\end{equation}
where $I_0$ is the input light power at the beamsplitter enhanced by
the PR gain and $I_\mathrm{SQL}$ is the light power needed by a
conventional interferometer (with no signal cavity)
to reach the SQL at sideband frequency
$\Omega=\gamma$.
$m$ is the mass of each arm cavity mirror, and $\omega_0$
is the carrier angular frequency. $h_\mathrm{SQL}$ in Eq.(\ref{eq:tf})
is the SQL for gravitational wave strain measurement, given by
$h_\mathrm{SQL}=\sqrt{8\hbar/m\Omega^2L^2}$.
%\begin{equation}
%h_\mathrm{SQL}=\sqrt{\frac{8\hbar}{m\Omega^2L^2}}.
%\end{equation}
Again, quantum noise is neglected in this
calculation of the predicted optical gain.

As mentioned above, 
the optical gain described by Eq.~(\ref{eq:tf})
exhibits a detuned RSE optical resonance peak
and an optical spring peak.
The shape of the optical gain as a function of frequency
is somewhat different than the (inverse of the) shape of the
quantum-limited sensitivity curve;
owing to the assumption that $h \gg a_i$,
quantum noise correlations will not be evident when the optical
gain is measured.

Several small-scale (``table-top'') experiments
have been used to study optical configurations
similar to that planned for Advanced LIGO,
and developed prototypes for the control topology 
required to operate them~\cite{Muller03,Mason03,Shaddock03,Miyakawa02}. 
The results of these experiments formed the basis for the 
Advanced LIGO design~\cite{Strain03}. 
More recently, 
Somiya {\it et al.}~\cite{Somiya05} developed and operated
a detuned RSE interferometer with suspended mirrors.
Optical springs have been observed
in detuned single Fabry-Perot cavities 
with low input power (no power recycling) and light masses, by
Sheard {\it et al.}~\cite{Sheard04} 
and Corbitt {\it et al.}~\cite{Corbitt05}.
The parametric instability in high
power stored cavities has also been explored~\cite{Corbitt05}.
\iffalse
{\bf LIGO folks: please point us to other relevant references
involving experiments in the AdLIGO-like configuration,
or experiment or theory with optical springs.}
\fi

%\section{Experimental setup}

\begin{figure}[tbp]
\includegraphics[width=8.5cm]{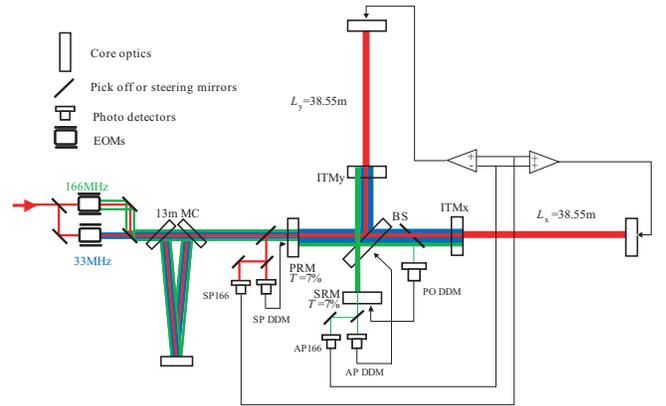}
\caption{\label{fig:40m_config}
Schematic diagram of the experimental setup of the 40m
interferometer. The optical configuration,
a Michelson interferometer with Fabry-Perot arms,
power recycling, and detuned RSE,
is similar to Advanced LIGO configuration.}
\end{figure}

The Caltech 40\,meter\,(40m) prototype interferometer 
was originally 
developed as a test bed for the initial LIGO optical configuration
and control system, and currently
it is used as a test bed for Advanced LIGO~\cite{Weinstein02,Miyakawa04}. 
The optical configuration of the 40m (Fig.~\ref{fig:40m_config})
is chosen to be similar to the optical configuration envisioned for 
Advanced LIGO.
The arm cavities were chosen to have the same finesse as 
Advanced LIGO (around 1200); the input test mass
mirrors (ITMs) have power transmission of 0.5\%.
The power recycling gain is designed to be 15,
and a signal recycling cavity detuning is chosen
to increase the detector bandwidth.

The light source is a 10\,W continuous Nd:YAG laser which has a frequency
stabilization servo, a pre-mode cleaner, and an intensity
stabilization servo.
Phase modulated RF sidebands are placed on 
the input beam at 33.2\,MHz and 166.0\,MHz
using electro-optic modulators in a Mach-Zehnder interferometer.
The light is attenuated, and 1\,W is injected
into a 13\,meter mode cleaner (MC) which consists of 3 suspended mirrors
forming a triangular cavity with 13\,m half-length.
This mode cleaner serves to further stabilize the laser frequency,
while transmitting the carrier light an both pairs of RF sidebands.
The beam passes through a Faraday isolator,
reflective mode-matching telescope,
and PZT-actuated mirrors
which steer the beam into the main interferometer.

All ten core optics (three for the mode cleaner and 
seven for the main interferometer) are
suspended as single pendula;
they behave like free masses above the pendulum resonant
frequency (around 0.8 Hz) and thus respond easily
to the optical spring
(the optomechanical rigidity is much larger than the
mechanical rigidity of the mirror suspension).
The suspended optics are placed on 
passive seismic isolation stacks,
within a single vacuum volume.
(In Advanced LIGO, multiple pendulum suspensions
and active seismic isolation systems will be used).

The main interferometer has 
five degrees of freedom that require length control:
the common and differential modes of the two Fabry-Perot arm cavities, 
the Michelson fringe, the PR cavity and the detuned SR cavity.
Output beams are monitored with RF photodiodes,
and length sensing signals are derived from demodulations
at 33.2, 132.8, 166.0 and 199.2 MHz.
Length control servos are implemented 
in a digital system 
to allow dynamical reconfiguration of the control topology 
and the signal filtering during and after lock acquisition.
The typical control bandwidth of these servos is 300 Hz.

Because of the complexity of the optical configuration
and the coupling of all the RF sidebands in the
detuned signal cavity,
lock acquisition and control of the interferometer is
far more challenging than in Initial LIGO.
Full lock acquisition and control in the desired configuration
was first achieved in November 2005,
through a process that will be described in a later publication.
The buildup of carrier and RF sideband fields in the
interferometer were then observed to be qualitatively as expected.
Arm cavity losses were somewhat higher than expected, however.
The achieved power recycling gain is 
about 5 and the arm cavity finesse is about 1200. 
The total power inside each arm is about 1.9\,kW.

%\section{Measurement}

In order to measure the optical response of the interferometer
to differential arm length changes (such as would arise
in the presence of a gravitational wave),
external sinusoidal forces are applied to the suspended optics
at the ends of the arms via magnetic actuation,
through the servo loop controlling that length degree of freedom.
The error signal for that servo loop
is extracted at the asymmetric port of the detector,
after transmission through the signal cavity.
The light is detected at the signal port, demodulated at 166\,MHz,
whitened, digitized, filtered, and then fed back to the
differential displacement of the two suspended optics 
at the ends of the arm cavities.

The optical gain of the differential mode of the arms is measured
as the spectral transfer function from the in-loop feedback signal to the
error signal. This transfer function includes,
in addition to the desired optical response,
the actuator pendulum
transfer function which has a $f^{-2}$ behavior above the 
pendulum resonant frequency of 0.8\,Hz.
There are also known whitening and anti-aliasing filters,
and time and phase delays associated with the
conversion from analog to digital and from digital to analog. 
The delays are measured
with another simpler optical configuration consisting of
a single Fabry-Perot arm cavity, and compensated. 

\begin{figure}[!tbhp]
\includegraphics[width=9cm]{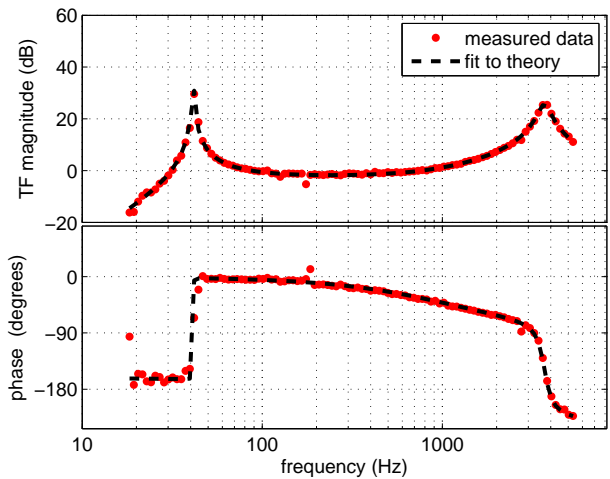}
\caption{\label{fig:os} 
The magnitude (top) and phase (bottom) response of 
the 40m interferometer to a swept-sine excitation of the 
arm length differential mode.
The points show the measured response,
while the dashed line is the prediction.
The peak at 41 Hz is due to the optical spring resonance
and the peak at around 3600 Hz is due to the 
optical resonance of the signal sideband in the RSE signal cavity.
}
\end{figure}

Fig~\ref{fig:os} shows the 
measured optical gain (solid line) and 
the prediction from Eq.~(\ref{eq:tf}) (dashed line).
In Fig~\ref{fig:os},
the peak at 41 Hz is due to the optical spring resonance
and the peak at approximately 3600 Hz is due to the 
optical resonance of the signal sideband in the RSE signal cavity.
Table~\ref{tab:table1} shows the
parameters which are used to calculate the model of the optical gain
for the differential mode of the arms. 

\begin{table}[!tbhp]
\caption{\label{tab:table1}40m parameters.}
\begin{ruledtabular}
\begin{tabular}{cc}
Quantity & Symbol \& value\\
\hline
Power at beam splitter & $I_0=4.2$\,W\\
Laser angular frequency & $\omega_0=1.8\times 10^{15}\,\mathrm{sec^{-1}}$\\
End-mirror mass & $m=1.276$\,kg\\
SR mirror transmissivity & $\tau=\sqrt{0.07}$ (amplitude)\\
Laser power to reach SQL & $I_\mathrm{SQL}=2.3$\,kW\\
Arm cavity half bandwidth & $\gamma=Tc/4L=2\pi\times1550\,\mathrm{sec^{-1}}$\\
Arm cavity length & $L=38.55$\,m\\
Input test mass transmissivity & $T=0.005$ (power)\\
SR cavity detuning & $\phi=\pi/2-0.39$\,rad\\
Homodyne phase & $\zeta=0.22$\,rad\\
\end{tabular}
\end{ruledtabular}
\end{table}

Degeneracies of parameters which determine the optical
response of the interferometer were investigated in~\cite{Buonanno03}.  
Of particular note are equations (13) of~\cite{Buonanno03}, which shows the
relation between the ITM transmittance, the signal mirror reflectivity, 
and the signal cavity detuning phase 
in determining the free RSE optical resonant frequency
and equation (49) which relates these
quantities, along with the circulating power, to the ponderomotive
rigidity (the optical spring peak).  
In producing Fig~\ref{fig:os}, we assume the
ITM and signal mirror transmittance to be at their design values, 
and vary the
detune phase, signal cavity losses, circulating laser power, and
measurement quadrature to fit the theoretical curve to the measured
data.  The values thus obtained are consistent with our expectations
based on the interferometer design, and on other
measurements made with the interferometer.

The quadrature $\zeta$ in Eq.~(\ref{eq:tf})
can be chosen by changing the RF demodulation
phase of the 166\,MHz local oscillator because the upper +166\,MHz
sideband is designed to be resonant in the combined
signal and power cavities,
while the lower -166\,MHz sideband is not resonant there.
This unbalanced RF
sideband at the detection port makes it possible to choose the
quadrature $\zeta$~\cite{Somiya03}. In this measurement, $\zeta$
is determined to be 0.22\,rad by fitting the measured response
in Fig~\ref{fig:os}.

The Michelson asymmetry combined with the detuned signal cavity causes
the control RF sidebands to be imbalanced in the interferometer.
This leads
to demodulation phase dependent offsets in the error signals derived
from these sidebands.  
These offsets are largely indistinguishable
from actual length deviations, and are strongly mixed among the short
degrees of freedom.  A lack of secondary diagnostics to precisely
determine the offset in the signal cavity length sensing means that the
signal cavity detuning was not known, operationally, to
a precision greater than a few percent.  Thus, we treat the detuning
here as a partially free parameter, and the exact detuning used to
produce the theoretical curve in Fig~\ref{fig:os} 
was determined through a fit to the data.  

\begin{figure}[!tbhp]
\includegraphics[width=8.2cm]{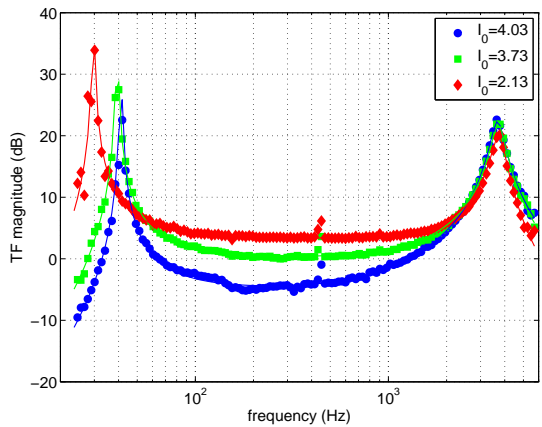}
\caption{\label{fig:TFpowers} 
The magnitude response of 
the 40m interferometer to a swept-sine excitation of the 
arm length differential mode, for three different values of
the effective incident beam power $I_0$.
The circular, square and triangular points (color online)
show the measured responses,
while the solid lines are the prediction
from Eq.~(\ref{eq:tf}).
}
\end{figure}

The frequency of the optical spring peak is a function
of the input power, as can be verified by varying the 
power incident on the beam splitter, $I_0$.
We can vary this power at the laser source, or 
equivalently, we can offset the common mode of the 
arm cavities from full resonance.
This method also introduces an optical spring 
effect in the common mode of the arms, which we observe to be in 
quantitative agreement with expectations.
In this way, we measure the optical gain in the differential mode 
with several different values of $I_0$, shown in Fig.~\ref{fig:TFpowers}.
Both the magnitude and phase (not shown) of the optical gain follow
the prediction from Eq.~(\ref{eq:tf}), with essentially no free parameters.
In particular, the dependence of the quadrature $\zeta$ 
on the offset of the common mode follows from a detailed numerical calculation.

The form of the measured optical gain and its dependence
on the effective laser power
are in good agreement with the
prediction, thus building confidence in both the 
theory and practical implementation of the 
RSE interferometer's response to differential displacement.
In particular, the theoretical prediction for 
the quantum-limited sensitivity achievable by Advanced LIGO
is supported. 
However, we do not report on the measured noise spectrum
or its comparison with prediction in this paper.
Other noise sources make it impossible at present
to achieve quantum-limited sensitivity
in the frequency range of interest for GW detection
with the 40m interferometer.

Note that the 180$^\circ$ phase advance in the optical spring resonance
results in an instability~\cite{Buonanno02}, but 
in practice it was not a problem to acquire
operational lock for this experiment because the control bandwidth was
about 300\,Hz,
well above the optical spring resonance at 41\,Hz.
% and the length control 
% servo has enough gain to suppress the instability of the optical spring. 
The servo unity gain frequency
in this measurement lies between the optical spring resonance
and the RSE optical resonance, in a region where the
phase of the optical gain is fairly flat.
This will probably not be the case in Advanced LIGO and other
next-generation GW detectors, and the stability of the
control servos must be carefully considered in the design.

%\begin{acknowledgments}
This work is supported by the National Science Foundation cooperative
agreement PHY0107417. This document has been assigned LIGO Laboratory
document number LIGO-P060007-01-R. We thank the many members of the
LIGO Laboratory, the LIGO Scientific Collaboration, 
% the engineering team 
and many valuable visitors to the 40 meter lab,
for their invaluable contributions.
%\end{acknowledgments}

%\newpage %Just because of unusual number of tables stacked at end

\end{document}